\documentclass[preprintnumbers, floatfix, onecolumn,
preprintnumbers, letterpaper, superscriptaddress,nofootinbib]{revtex4}
\usepackage{makeidx}
\usepackage{amsfonts}
\usepackage{amsmath}
\usepackage{amssymb,epsf}
\usepackage{latexsym}
\usepackage{graphicx,epsfig}
\usepackage{amssymb}
\usepackage{subfigure}
\usepackage[colorlinks=true,citecolor=blue,linkcolor=blue,urlcolor=black]{hyperref}

\begin{document}

\title{Thermodynamics of Gauss-Bonnet-dilaton Lifshitz black branes}
\author{M. Kord Zangeneh}
\email{mkzangeneh@shirazu.ac.ir}
\affiliation{Physics Department and Biruni Observatory, Shiraz University, Shiraz 71454,
Iran}
\author{M. H. Dehghani}
\email{mhd@shirazu.ac.ir}
\affiliation{Physics Department and Biruni Observatory, Shiraz University, Shiraz 71454,
Iran}
\affiliation{Research Institute for Astronomy and Astrophysics of Maragha (RIAAM), P.O.
Box 55134-441, Maragha, Iran}
\author{A. Sheykhi}
\email{asheykhi@shirazu.ac.ir}
\affiliation{Physics Department and Biruni Observatory, Shiraz University, Shiraz 71454,
Iran}
\affiliation{Research Institute for Astronomy and Astrophysics of Maragha (RIAAM), P.O.
Box 55134-441, Maragha, Iran}

\begin{abstract}
We explore an effective supergravity action in the presence of a massless
gauge field which contains the Gauss-Bonnet term as well as a dilaton field. We
construct a new class of black brane solutions of this theory with the Lifshitz
asymptotic by fixing the parameters of the model such that the asymptotic
Lifshitz behavior can be supported. Then we construct the well-defined
finite action through the use of the counterterm method. We also obtain two
independent constants along the radial coordinate by combining the equations
of motion. Calculations of these two constants at infinity through the use
of the large-$r$ behavior of the metric functions show that our solution
respects the no-hair theorem. Furthermore, we combine these two constants in
order to get a constant $\mathcal{C}$ which is proportional to the energy of
the black brane. We calculate this constant at the horizon in terms of the
temperature and entropy, and at large-$r$ in terms of the geometrical mass.
By calculating the value of the energy density through the use of the counterterm
method, we obtain the relation between the energy density, the temperature,
and the entropy. This relation is the generalization of the well-known Smarr
formula for AdS black holes. Finally, we study the thermal stability of our
black brane solution and show that it is stable under thermal perturbations.
\end{abstract}

\maketitle

\section{INTRODUCTION}

The formulation of the quantum theory of gravity and its applications to
physical systems in order to understand physics at strong gravity is one of
the most interesting challenges of modern theoretical physics. The area where
quantum gravity may play a significant role includes cosmology and black
hole physics. Although the leading candidates are the ten-dimensional
superstring theories, most analyses have been performed by using low-energy
effective theories inspired by string theories. This is due to the fact that
it is difficult to study geometrical settings in superstring theories. The
effective theories are the supergravities which typically involve not only
the metric but also a dilaton field (as well as several gauge fields). On
the other hand, the effective supergravity action coming from superstrings
contains higher-order curvature correction terms. The simplest correction is
the Gauss-Bonnet (GB) term coupled to the dilaton field in the low-energy
effective heterotic string \cite{Str}. It is then natural to ask how the
black hole solutions are affected by higher-order terms in these effective
theories. To our knowledge, just one exact solution of such a
theory has been obtained \cite{Char}. Indeed in Ref. \cite{Char}, a
dilatonic Einstein-Gauss-Bonnet (EGB) theory with a nonminimal coupling
between the EGB term and dilaton field has resulted by applying the general
Kaluza-Klein reduction on EGB gravity and an exact solution has been
introduced. Also in Ref. \cite{Ohta1}, it was shown that the anti-de Sitter
(AdS) metric can be an exact solution of this theory. In addition,
asymptotically AdS solutions of this theory have been considered in Refs.
\cite{Ohta1,Ohta2} numerically. Here, we would like to find Lifshitz
solutions of this theory. The motivation for this investigation is that
Lifshitz black holes has received much attention recently. Indeed, black
hole configurations in Lifshitz spacetime are dual to nonrelativistic
conformal field theories enjoying anisotropic conformal transformations
\begin{equation}
t\rightarrow \lambda ^{z}t,\text{ \ \ \ \ \ \ \ \ \ }\vec{\mathbf{x}}%
\rightarrow \lambda \vec{\mathbf{x}},  \label{aniso}
\end{equation}%
where the constant $z>1$ called the dynamical exponent and shows the
anisotropy between space and time. In other words, while AdS black holes are
dual to scale invariant relativistic field theories which respect the
isotropic conformal transformation \cite{Mald}%
\begin{equation}
t\rightarrow \lambda t,\text{ \ \ \ \ \ \ \ \ \ }\vec{\mathbf{x}}\rightarrow
\lambda \vec{\mathbf{x}},  \label{isot}
\end{equation}%
Lifshitz black holes are dual to nonrelativistic field theories.
Furthermore, there are many situations where isotropic conformal
transformation (\ref{isot}) is not respected. For instance in many condensed
matter systems, there are phase transitions governed by fixed points which
exhibit dynamical scaling (\ref{aniso}). The gravity models dual to such
systems are no longer AdS and one needs a spacetime that its boundary
respects anisotropic conformal transformation (\ref{aniso}). This spacetime,
which is known as Lifshitz spacetime, was first introduced in \cite{Lif} as
\begin{equation}
ds^{2}=-\frac{r^{2z}}{l^{2z}}dt^{2}+\frac{l^{2}dr^{2}}{r^{2}}+r^{2}d\vec{%
\mathbf{x}}^{2}.
\end{equation}

From the beginning of the introduction of the Lifshitz spacetime, it was known
that this is not a vacuum solution of Einstein gravity or even the Einstein
equation with the cosmological constant in the case of an arbitrary value of $z$%
. Therefore one needs some matter sources or higher-curvature corrections in
order to guarantee that the asymptotic behavior of spacetime is Lifshitz. One of
the matter sources considered in many works is a massive gauge field \cite%
{manal1,manal2,cmet,cmlt,cmlnt}. In the latter case, some exact solutions
have been obtained more often for fixed $z$ \cite{manal1,manal2}. However,
it seems impossible in some cases to find an exact solution of the field
equations of motion especially for arbitrary $z$, either in Einstein \cite%
{manal1,cmet} or in Lovelock gravities \cite{cmlt,cmlnt}. But, it is
possible to study the thermodynamics of these black hole configurations by using
a conserved quantity along the radial coordinate \cite{cmet,cmlt}. Another
way to guarantee the Lifshitz asymptotic behavior is by considering higher
curvature corrections \cite{hcc}. In addition, asymptotic supporting matter
source can also be chosen to be a dilaton field and a massless gauge field
\cite{dcanal,Zang,peet}. One of the advantages of the latter matter source
over the massive gauge field is that in this case it is possible to find
analytic Lifshitz black hole solutions in Einstein gravity for arbitrary $z$%
. Some charged exact solutions have been presented in \cite{dcanal,Zang}. In
\cite{peet}, the thermal behavior of uncharged dilaton Lifshitz black branes
has been studied by using perturbation theory.

Motivated by the above two paragraphs, it is interesting to study the
thermodynamics of Lifshitz black brane solutions in the effective
supergravity action coming from superstrings, which contains higher-order
curvature correction terms and a dilaton field. In this paper we attack 
this problem. We shall investigate thermodynamics of Lifshitz black branes
in the presence of a dilaton, a massless gauge field and GB term. Although
exact dilatonic Lifshitz solutions have been introduced in Einstein gravity
\cite{dcanal,Zang}, no exact asymptotically Lifshitz solution has been
obtained in dilaton EGB gravity. Therefore, we seek the thermodynamics of
Lifshtiz black branes in GB-dilaton gravity using a conserved quantity along
the radial coordinate as in the case of massive gauge field matter sources.
We find the relation between energy density, temperature and entropy by
using the fact that the values of our constant quantity along the radial
coordinate are related to the temperature and entropy at the horizon and to
the energy density for large $r$.

The layout of this paper is as follows. In Sec. \ref{Field}, we obtain the
one-dimensional Lagrangian and derive the field equations of motion for a
general spherically symmetric spacetime. In Sec. \ref{Lif}, we show that
this theory can accept the Lifshitz metric as its solution. Two constants along
the radial coordinate will be introduced in Sec. \ref{Cons}. Section \ref%
{Finite} is devoted to the generalization of the counterterm method of Ref.
\cite{Ross} in the presence of a dilaton field. In Sec. \ref{Therm}, we
calculate the thermodynamic and conserved quantities of asymptotic
Lifshitz black branes and obtain the Smarr relation. We also show that our
solution is thermally stable. We finish our paper with summary and some
concluding remarks.

\section{FIELD EQUATIONS\label{Field}}

The effective action of the heterotic string theory in the Einstein frame in the
presence of a gauge field may be written as
\begin{equation}
I_{\mathrm{bulk}}=\frac{1}{16\pi }\int_{\mathcal{M}}d^{n+1}x\sqrt{-g}\left(
R-2\Lambda +\alpha e^{\eta \Phi }\mathcal{L}_{2}-\frac{4}{n-1}\left(
\partial \Phi \right) ^{2}-e^{-4\lambda \Phi /(n-1)}F\right) ,  \label{Act1}
\end{equation}%
where $F=F_{\mu \nu }F^{\mu \nu }$, $F_{\mu \nu }=\partial _{\lbrack \mu
}U_{\nu ]}$, $\eta $ is a constant, $\lambda $ is the coupling constant of
dilaton and matter, $\Lambda $ is the cosmological constant, $\alpha $ is
the Gauss-Bonnet (GB) coefficient, $\mathcal{L}_{2}=R_{\mu \nu \gamma \delta
}R^{\mu \nu \gamma \delta }-4R_{\mu \nu }R^{\mu \nu }+R^{2}$ is the GB
Lagrangian and $\Phi $ is the dilaton field. While the ten-dimensional
critical string theory predicts the coupling constant $\eta \neq 0$, we set
it equal to zero. This is due to the fact that, as we will see later and
also in the absence of the Gauss-Bonnet term considered in Ref. \cite{Zang}, the
dilaton goes to infinity as $r$ goes to infinity. Thus, $e^{\eta \Phi }$
becomes very large at large $r$ which effectively rescales the coupling
constant $\alpha $ to large values. Indeed, a large value $e^{\eta \Phi }$
leads to a large modification to general relativity which is ruled out by
the weak field approximation \cite{Ayz}. So, we set $\eta =0$. In this case the
GB term for $n\leq 3$ does not contribute to the field equations and is a
total derivative in the action. Here we consider the GB coefficient $\alpha $
positive as in the heterotic string theory \cite{STR}.

We write the spherically symmetric gauge fields and the metric of an $(n+1)$%
-dimensional asymptotically Lifshitz static spacetime with zero curvature
boundary as
\begin{eqnarray}
U &=&qe^{K(r)}dt,  \label{Ai} \\
ds^{2} &=&-e^{2A(r)}dt^{2}+e^{2C(r)}dr^{2}+l^{2}e^{2B(r)}d\vec{\mathbf{x}}%
^{2},  \label{metric}
\end{eqnarray}%
where $d\vec{\mathbf{x}}^{2}=\sum\limits_{i=2}^{n}(dx^{i})^{2}$. Inserting (%
\ref{Ai}) and (\ref{metric}) into the action (\ref{Act1}) and integrating by
part, one obtains the one-dimensional Lagrangian as $\mathcal{L}_{1D}=\left(
n-1\right) l^{n-1}\mathcal{\tilde{L}}_{1D}$ where
\begin{eqnarray}
\mathcal{\tilde{L}}_{1D} &=&-\frac{2\Lambda }{n-1}%
e^{A+C+(n-1)B}+e^{A-C+(n-1)B}\left[ 2A^{\prime }B^{\prime }+(n-2)B^{\prime 2}%
\right]  \notag \\
&&-2\tilde{\alpha}e^{A-3C+(n-1)B}\left[ \frac{2}{3}B^{\prime 3}A^{\prime }+%
\frac{n-4}{6}B^{\prime 4}\right] -\frac{4}{(n-1)^{2}}e^{A-C+(n-1)B}\Phi
^{\prime 2}+\frac{2q^{2}}{n-1}e^{-A-C+(n-1)B-4\lambda \Phi
/(n-1)+2K}K^{\prime 2},  \notag \\
&&  \label{L1D}
\end{eqnarray}%
$\tilde{\alpha}=\alpha (n-2)(n-3)$ and prime denotes the derivative
with respect to the $r$ coordinate. Varying the action (\ref{L1D}) with respect
to $A(r)$, $B(r)$, $C(r)$, $\Phi (r)$, and $K(r)$, respectively, one obtains
the following equations of motion:
\begin{eqnarray}
E_{1} &=&e^{A-C+(n-1)B}\left[ B^{\prime \prime }+\frac{n}{2}B^{\prime
2}-B^{\prime }C^{\prime }+\frac{2}{(n-1)^{2}}\Phi ^{\prime 2}\right] -\frac{%
\tilde{\alpha}n}{2}e^{A-3C+(n-1)B}\left[ \frac{4}{n}B^{\prime \prime
}B^{\prime 2}+B^{\prime 4}-\frac{4}{n}B^{\prime 3}C^{\prime }\right]  \notag
\\
&&+\frac{\Lambda }{n-1}e^{A+C+(n-1)B}+\frac{q^{2}}{n-1}e^{-A-C+(n-1)B-4%
\lambda \Phi /(n-1)+2K}K^{\prime 2}=0,  \label{EOM1}
\end{eqnarray}%
{\small
\begin{eqnarray}
E_{2} &=&-e^{A-C+(n-1)B}\left[ -(n-2)B^{\prime \prime }-A^{\prime \prime }-%
\frac{(n-1)(n-2)}{2}B^{\prime 2}-A^{\prime 2}+(n-2)B^{\prime }(C^{\prime
}-A^{\prime })+A^{\prime }C^{\prime }-\frac{2}{n-1}\Phi ^{\prime 2}\right]
\notag \\
&&-2\tilde{\alpha}e^{A-3C+(n-1)B}\left\{ B^{\prime \prime }B^{\prime }\left[
2A^{\prime }+(n-4)B^{\prime }\right] +B^{\prime 2}\left[ A^{\prime \prime }+%
\frac{(n-2)(n-3)-2}{4}B^{\prime 2}+B^{\prime }\left( (n-2)A^{\prime
}-(n-4)C^{\prime }\right) +A^{\prime }\left( A^{\prime }-3C^{\prime }\right) %
\right] \right\}  \notag \\
&&+e^{A+C+(n-1)B}\Lambda -q^{2}e^{-A-C+(n-1)B-4\lambda \Phi
/(n-1)+2K}K^{\prime 2}=0,  \label{EOM2}
\end{eqnarray}%
}
\begin{eqnarray}
E_{3} &=&e^{A-C+(n-1)B}\left[ A^{\prime }B^{\prime }+\frac{(n-2)}{2}%
B^{\prime 2}-\frac{2}{(n-1)^{2}}\Phi ^{\prime 2}\right] -2\tilde{\alpha}%
e^{A-3C+(n-1)B}\left[ A^{\prime }B^{\prime 3}+\frac{n-4}{4}B^{\prime 4}%
\right]  \notag \\
&&+\frac{\Lambda }{n-1}e^{A+C+(n-1)B}+\frac{q^{2}}{n-1}e^{-A-C+(n-1)B-4%
\lambda \Phi /(n-1)+2K}K^{\prime 2}=0,  \label{EOM3}
\end{eqnarray}

\begin{equation}
E_{4}=\frac{4}{(n-1)^{2}}\left\{ -e^{A-C+(n-1)B}\left[ \Phi ^{\prime \prime
}+\Phi ^{\prime }\left( A^{\prime }-C^{\prime }+(n-1)B^{\prime }\right) %
\right] +q^{2}\lambda e^{-A-C+(n-1)B-4\lambda \Phi /(n-1)+2K}K^{\prime
2}\right\} =0,  \label{EOM4}
\end{equation}

\begin{equation}
E_{5}=\frac{8q^{2}}{(n-1)^{2}}e^{-A-C+(n-1)B-4\lambda \Phi /(n-1)+2K}\left\{
K^{\prime \prime }+K^{\prime 2}+K^{\prime }\left[ -\frac{4}{n-1}\Phi
^{\prime }-A^{\prime }-C^{\prime }+(n-1)B^{\prime }\right] \right\} =0.
\label{EOM6}
\end{equation}%
One should also note that there is a relation between the equations of motion,

\begin{equation}
A^{\prime }E_{1}+B^{\prime }E_{2}+C^{\prime }E_{3}-E_{3}^{\prime }+\Phi
^{\prime }E_{4}+K^{\prime }E_{5}=0,  \label{relat}
\end{equation}%
where prime denotes the derivative with respect to $r$. The above equation (%
\ref{relat}) reduces the number of independent field equations to four. One
may note that $E_{5}$ is the Maxwell equation that can be solved for $K(r)$ as:

\begin{equation}
\left( e^{K(r)}\right) ^{\prime }=l^{-n}e^{4\lambda \Phi /(n-1)+A+C-(n-1)B},
\label{K'}
\end{equation}%
and, therefore, we leave with three independent equations of motion.

\section{LIFSHITZ SOLUTION \label{Lif}}

We used the metric (\ref{metric}) in order to find the constants of the
system along the radial coordinate $r$, which will be done in the next
section. Here we want to find the Lifshitz solutions. In order to do this, we
use the standard form of the asymptotic Lifshitz metric through the use of the
following transformations:%
\begin{eqnarray}
A(r) &=&\frac{1}{2}\ln \left( \frac{r^{2z}}{l^{2z}}f(r)\right) ,  \notag \\
C(r) &=&-\frac{1}{2}\ln \left( \frac{r^{2}}{l^{2}}g(r)\right) ,  \notag \\
B(r) &=&\ln \frac{r}{l},  \notag \\
K(r) &=&\ln \frac{k(r)}{l^{z}}.  \label{trans}
\end{eqnarray}
Using the above transformation (\ref{trans}), the metric and gauge field can
be written as%
\begin{eqnarray*}
ds^{2} &=&-\frac{r^{2z}}{l^{2z}}f(r)dt^{2}+\frac{l^{2}dr^{2}}{r^{2}g(r)}%
+r^{2}d\vec{\mathbf{x}}^{2}, \\
U &=&\frac{q}{l^{z}}k(r)dt.
\end{eqnarray*}%
In this frame, the solution of the electromagnetic equation is:
\begin{equation}
k^{\prime }(r)=e^{4\lambda \Phi /(n-1)}r^{z-n}\sqrt{\frac{f}{g}}.  \label{k'}
\end{equation}%
Here we choose the horizon as the reference point of the potential. Thus,
one obtains%
\begin{equation}
k(r)=\int^{r}e^{4\lambda \Phi /(n-1)}r^{z-n}\sqrt{\frac{f}{g}}dr+D,
\label{ki}
\end{equation}%
with
\begin{equation*}
D=-\int^{r_{0}}e^{4\lambda \Phi /(n-1)}r^{z-n}\sqrt{\frac{f}{g}}dr,
\end{equation*}%
where $r_{0}$ is the horizon radius. We first investigate the possibility of
having ($n+1$)-dimensional Lifshitz solutions. For this aim, the field
equations should be satisfied for $f(r)=g(r)=1$. In this case, by using (\ref%
{trans}) and (\ref{ki}), Eqs. (\ref{EOM1})-(\ref{EOM4}) reduce to

\begin{equation}
-r^{2}\Phi _{lif}^{\prime 2}-\frac{n(n-1)^{2}}{4}+\frac{\tilde{\alpha}%
n(n-1)^{2}}{4l^{2}}-\frac{n-1}{2}\left[ \Lambda l^{2}+\frac{q^{2}e^{4\lambda
\Phi _{lif}/(n-1)}}{r^{2(n-1)}}\right] =0  \label{E1}
\end{equation}
{\small
\begin{equation}
r^{2}\Phi _{lif}^{\prime 2}+(n-1)(n-2)\left[ \frac{(n-1)}{4}+\frac{z^{2}}{%
2(n-2)}+\frac{z}{2}\right] -\frac{\tilde{\alpha}(n-1)}{l^{2}}\left[ \frac{%
(n-1)(n-4)}{4}+(n-2)z+z^{2}\right] +\frac{n-1}{2}\left[ \Lambda l^{2}-\frac{%
q^{2}e^{4\lambda \Phi _{lif}/(n-1)}}{r^{2(n-1)}}\right] =0,
\end{equation}
}
\begin{equation}
r^{2}\Phi _{lif}^{\prime 2}-\frac{(n-1)^{2}\left( (n-2)+2z\right) }{4}+\frac{%
\tilde{\alpha}(n-1)^{2}}{l^{2}}\left[ z+\frac{n-4}{4}\right] -\frac{n-1}{2}%
\left[ \Lambda l^{2}+\frac{q^{2}e^{4\lambda \Phi _{lif}/(n-1)}}{r^{2(n-1)}}%
\right] =0,  \label{E3}
\end{equation}

\begin{equation}
r\Phi _{lif}^{\prime \prime }+(z+n)\Phi _{lif}^{\prime }-\frac{\lambda
q^{2}e^{4\lambda \Phi _{lif}/(n-1)}}{r^{2(n-1)+1}}=0.  \label{E4}
\end{equation}%
Subtracting (\ref{E1}) from (\ref{E3}) one can find

\begin{equation}
2r^{2}\Phi _{lif}^{\prime 2}-\frac{(n-1)^{2}\left( z-1\right) }{2}\left( 1-%
\frac{2\tilde{\alpha}}{l^{2}}\right) =0,
\end{equation}%
with the solution
\begin{gather}
\Phi _{lif}(r)=\xi \ln \left( \frac{r}{b}\right) ,  \label{Phi} \\
\xi =\frac{n-1}{2l}\sqrt{\left( z-1\right) \left( l^{2}-2\tilde{\alpha}%
\right) }.  \notag
\end{gather}%
Substituting the solution (\ref{Phi}) in the Eqs. (\ref{E1})-(\ref{E4}), it
is a matter of calculation to show that they are fully satisfied, provided
\begin{gather}
\lambda =\frac{(n-1)l}{\sqrt{\left( z-1\right) \left( l^{2}-2\tilde{\alpha}%
\right) }},  \notag \\
q=\frac{b^{n-1}}{\sqrt{2}l}\sqrt{\left( z-1\right) \left( l^{2}-2\tilde{%
\alpha}\right) \left( z+n-1\right) },  \notag \\
\Lambda =-\frac{(z+n-1)(z+n-2)}{2l^{2}}+\frac{\tilde{\alpha}\left[
2z^{2}+2\left( 2n-3\right) z+(n-2)(n-3)-2\right] }{2l^{4}}.
\label{Constants}
\end{gather}%
Please note that in the case of $\alpha =0$ and $z=1$, $\Lambda $ reduces to
$-n(n-1)/2l^{2}$ and in the case of $z=1$, it reduces to $-n(n-1)\left( 1-%
\tilde{\alpha}/l^{2}\right) /2l^{2}$\ as expected in the cases of AdS
spacetimes in Einstein and Einstein-Gauss-Bonnet gravity, respectively. We
can also find the asymptotic value of $k(r)$ as $r$ goes to infinity by
using (\ref{EOM6}), (\ref{Phi}) and (\ref{Constants}) as
\begin{equation}
k_{lif}=\frac{b^{2-2n}r^{z+n-1}}{z+n-1}{.}
\end{equation}

\section{THE CONSTANT ALONG THE\ RADIAL\ COORDINATE \label{Cons}}

As in the case of the Einstein equation in the presence of a dilaton \cite{peet},
one can find two independent constants along the radial coordinate $r$ for
this spacetime. It could be checked that there are two combinations of field
equations which are exact differentials: {\small
\begin{equation}
E_{1}-\frac{E_{2}}{n-1}+E_{5}=-\frac{1}{n-1}\left[ e^{A-C+(n-1)B}\left(
A^{\prime }-B^{\prime }\right) +2\tilde{\alpha}e^{A-3C+(n-1)B}\left(
B^{\prime 3}-A^{\prime }B^{\prime 2}\right) -2q^{2}e^{-A-C+(n-1)B-4\lambda
\Phi /(n-1)+2K}K^{\prime }\right] ^{\prime }=0.
\end{equation}
}

\begin{equation}
E_{4}+\frac{2\lambda E_{5}}{n-1}=-\frac{4}{(n-1)^{2}}\left[
e^{A-C+(n-1)B}\Phi ^{\prime }-\lambda q^{2}e^{-A-C+(n-1)B-4\lambda \Phi
/(n-1)+2K}K^{\prime }\right] ^{\prime }=0.
\end{equation}%
This fact results in two independent constants:

\begin{eqnarray}
\mathcal{C}_{1} &=&-\frac{1}{n-1}\left[ e^{A-C+(n-1)B}\left( A^{\prime
}-B^{\prime }\right) +2\tilde{\alpha}e^{A-3C+(n-1)B}\left( B^{\prime
3}-A^{\prime }B^{\prime 2}\right) -2q^{2}e^{-A-C+(n-1)B-4\lambda \Phi
/(n-1)+2K}K^{\prime }\right]  \notag \\
&=&-\frac{1}{2(n-1)l^{n+z}}\left\{ \left[ 1-\frac{2\tilde{\alpha}}{l^{2}}g%
\right] \left[ r^{z+n}f^{\prime }\sqrt{\frac{g}{f}}+2r^{z+n-1}\sqrt{fg}(z-1)%
\right] -4q^{2}k\right\} ,  \label{c1}
\end{eqnarray}

\begin{eqnarray}
\mathcal{C}_{2} &=&-\frac{4}{(n-1)^{2}}\left[ e^{A-C+(n-1)B}\Phi ^{\prime
}-\lambda q^{2}e^{-A-C+(n-1)B-4\lambda \Phi /(n-1)+2K}K^{\prime }\right]
\notag \\
&&-\frac{4}{(n-1)^{2}l^{n+z}}\,{\left[ {r}^{n+z}\sqrt{fg}\Phi ^{\prime
}-\lambda q^{2}k\right] },  \label{c2}
\end{eqnarray}%
We pause to remark that $q$ and $\Phi $ vanish for $z=1$, and the theory
reduces to Einstein-Gauss-Bonnet (EGB) gravity. Thus, with $f(r)=g(r)$ the
first constant (\ref{c1}) reduces to
\begin{equation*}
\mathcal{C}_{1}=\frac{r^{n+1}}{2(n-1)l^{n+1}}\left( f-\frac{\tilde{\alpha}}{%
l^{2}}f^{2}\right) ^{\prime },
\end{equation*}%
which is known to be constant in EGB gravity and is proportional to the mass
parameter of the spacetime. Also note that the second constant is zero in
EGB gravity.

Combining the constants (\ref{c1}) and (\ref{c2}), one can get a constant
which is very useful in our future discussions in this work:%
\begin{eqnarray}
\mathcal{C} &=&-2(n-1)l^{n-1}\left( \mathcal{C}_{1}-\frac{n-1}{2\lambda }%
\mathcal{C}_{2}\right)  \notag \\
&=&\frac{r^{n+z}}{l^{z+1}}\left\{ \left( 1-\frac{2\tilde{\alpha}}{l^{2}}%
g\right) \left[ f^{\prime }\sqrt{\frac{g}{f}}+2r^{-1}\sqrt{fg}(z-1)\right] \,%
{-\frac{4{\sqrt{fg}\Phi ^{\prime }}}{\lambda }}\right\} .  \label{c0}
\end{eqnarray}%
In this section, we want to calculate the constant $\mathcal{C}$, which is
conserved along the radial coordinate $r$. Since there is no exact
GB-Lifshitz-dilaton solution, we calculate it at the horizon and at
infinity. We will use this to relate the constant that appears in the
expansion at $r=\infty $ to the coefficients at the horizon.

\subsection{$\mathcal{C}$ at the horizon}

Considering nonextreme black branes, one can assume that $f(r)$ and $g(r)$
go to zero linearly at the horizon. Also, we have chosen the reference point
of $k(r)$ at the horizon. Thus, one can write
\begin{eqnarray}
f(r) &=&f_{1}\left\{
(r-r_{0})+f_{2}(r-r_{0})^{2}+f_{3}(r-r_{0})^{3}+f_{4}(r-r_{0})^{4}+...\right%
\} ,  \notag \\
g(r)
&=&g_{1}(r-r_{0})+g_{2}(r-r_{0})^{2}+g_{3}(r-r_{0})^{3}+g_{4}(r-r_{0})^{4}+...,
\notag \\
k(r)
&=&k_{1}(r-r_{0})+k_{2}(r-r_{0})^{2}+k_{3}(r-r_{0})^{3}+k_{4}(r-r_{0})^{4}+...,
\notag \\
\Phi (r) &=&\Phi _{0}+\Phi _{1}(r-r_{0})+\Phi _{2}(r-r_{0})^{2}+\Phi
_{3}(r-r_{0})^{3}+\Phi _{4}(r-r_{0})^{4}+....  \label{ExpH}
\end{eqnarray}%
One can solve for the various coefficients by inserting these expansions
into the equations of motion arising from the action (\ref{Act1}) for the
metric (\ref{metric}) with the conditions (\ref{Constants}).

The constant $\mathcal{C}$ (\ref{c0}) can be evaluated at $r=r_{0}$ by using
the above expansion. One obtains%
\begin{equation}
\mathcal{C}=\frac{r_{0}^{z+n}\sqrt{f_{1}g_{1}}}{l^{z+1}}.  \label{CH}
\end{equation}%
This must be preserved along the flow in $r$.

\subsection{$\mathcal{C}$ at infinity}

We now turn to the calculation of $\mathcal{C}$\ at large $r$. In order to
do this, we investigate the behavior of the metric functions at large $r$ by
using straightforward perturbation theory. Using the following expansions
\begin{eqnarray*}
f(r) &=&1+\varepsilon f_{1}(r), \\
g(r) &=&1+\varepsilon g_{1}(r), \\
\Phi (r) &=&\Phi _{lif}+\varepsilon \Phi _{1}(r),
\end{eqnarray*}%
and finding the field equations (\ref{EOM1}-\ref{EOM3}) up to the first
order in $\varepsilon $, we obtain
\begin{eqnarray}
0 &=&{\left( n-1\right) \left[ {\left( {l}^{2}-2\,\tilde{\alpha}\right)
rf_{1}^{\prime }+}\left( z+n-1\right) {l}^{2}{g_{1}}-2\tilde{\alpha}\,\left(
3z+\,n-3\right) {g_{1}}\right] }+{4{l}\sqrt{(z-1)({l}^{2}-2\tilde{\alpha})}%
\left[ \left( z+n-1\right) \Phi _{1}-r\Phi _{1}^{\prime }\right] }  \notag \\
0 &=&{\left( l^{2}-2\,\tilde{\alpha}\right) \left[ {r}^{2}f_{1}^{\prime
\prime }+\left( 2z+n-1\right) rf_{1}^{\prime }\right] }+\,{\left[ \left(
z+n-2\right) {l}^{2}-2\,\tilde{\alpha}\,\left( 3\,z+n-4\right) \right]
rg_{1}^{\prime }}+{\left( z+n-1\right) }  \notag \\
&&{\left[ \left( 2z+n\,-3\right) {l}^{2}-\,2\tilde{\alpha}\left(
4z+\,n-5\right) \right] g_{1}}-{4{l}\sqrt{(z-1)({l}^{2}-2\tilde{\alpha})}%
\left[ \left( z+n-1\right) \Phi _{1}+r\Phi _{1}^{\prime }\right] }  \notag \\
0 &=&{\left( n-1\right) \sqrt{(z-1)({l}^{2}-2\tilde{\alpha})}\left[ 2\left(
z+n-1\right) g_{1}+{r\left( f_{1}^{\prime }+g_{1}^{\prime }\right) }\right]
+4{l}}\left[ {{r}^{2}\Phi _{1}^{\prime \prime }+\left( z+n\right) r\Phi
_{1}^{\prime }-2\left( n-1\right) \left( z+n-1\right) \Phi _{1}}\right]
\notag \\
&&  \label{infinity}
\end{eqnarray}%
Demanding the fact that the solutions corresponding to these field equations
should go to zero as $r\rightarrow \infty $, one can find the desired
solutions of Eqs. (\ref{infinity}) as
\begin{eqnarray}
f_{1}(r) &=&-\frac{C_{1}}{r^{n+z-1}}+\frac{C_{2}}{r^{\left( n+z-1+\gamma
\right) /2}},  \notag \\
g_{1}(r) &=&-\frac{\left( {l}^{2}-2\tilde{\alpha}\right) \left( z+n-1\right)
\left( z+n-2\right) C_{1}}{\mathcal{G}r^{n+z-1}}-\frac{\mathcal{TK}_{+}C_{2}%
}{r^{\left( n+z-1+\gamma \right) /2}},  \notag \\
\Phi _{1}(r) &=&-\frac{\left( n-1\right) \left( z-1\right) ^{3/2}\tilde{%
\alpha}\sqrt{{l}^{2}-2\tilde{\alpha}}C_{1}}{2l\mathcal{G}r^{n+z-1}}+\frac{%
\left( n-1\right) \sqrt{{l}^{2}-2\tilde{\alpha}}\mathcal{TK}_{+}C_{2}}{4{l}%
\sqrt{z-1}r^{\left( n+z-1+\gamma \right) /2}}.  \label{Solinf}
\end{eqnarray}%
where%
\begin{gather*}
\gamma =\sqrt{{\left( z+n-1\right) \left( 9z+9n-17\right) +}\frac{16\tilde{%
\alpha}\left( z-1\right) ^{2}}{{l}^{2}-2\,\tilde{\alpha}}}, \\
\mathcal{G}=({l}^{2}-2\tilde{\alpha})\left( z+n-1\right) \left( z+n-2\right)
-4\tilde{\alpha}(z-1)(n-1), \\
\mathcal{K}_{\pm }=\left( {l}^{2}-2\tilde{\alpha}\right) \left( z+n-1\right)
\left( z+n-2\right) -\tilde{\alpha}\left( z-1\right) (n-z+1\pm \gamma ), \\
\mathcal{T}^{-1}=(l^{2}-2\tilde{\alpha})\left( z+n-1\right) \left(
z+n-2\right) -2\tilde{\alpha}\left( z-1\right) n-\frac{8\tilde{\alpha}%
^{2}\left( z-1\right) ^{2}}{(l^{2}-2\tilde{\alpha})}, \\
\mathcal{W}=({l}^{2}-2\tilde{\alpha})\left( n+z-2\right) \gamma -\tilde{%
\alpha}\left( z-1\right) \left( 3\gamma +9n+7z-15\right) .
\end{gather*}%
Substituting (\ref{Solinf}) in (\ref{ki}), one could find the large-$r$
behavior of $k(r)$ as
\begin{equation}
k(r)=k_{lif}-\varepsilon \frac{\mathcal{T}\left( \mathcal{W}+\mathcal{K}%
_{-}\right) C_{2}}{4{b}^{2(n-1)}\left( z-1\right) r^{\left( \gamma
-n-z+1\right) /2}}  \label{kinf}
\end{equation}%
It is easy to check that $(\gamma-n-z+1)/2$, which is the power
of $r$ in the denominator of (\ref{kinf}), is always positive and, therefore, $%
k(r)\rightarrow k_{lif}$ as $r\rightarrow \infty $.

Now we want to calculate the conserved quantity $\mathcal{C}$ given by (\ref%
{c0}). The constant $\mathcal{C}$ can be obtained as%
\begin{equation}
\mathcal{C}={\frac{\left( z+n-1\right) \left( l^{2}-2\tilde{\alpha}\right) %
\left[ \left( z+n-2\right) \left( z+n-1\right) \left( l^{2}-2\tilde{\alpha}%
\right) +2\tilde{\alpha}\left( z-1\right) ^{2}\right] C_{1}}{\mathcal{G}{l}%
^{z+3}}}.  \label{c}
\end{equation}%
For $\alpha =0$, $\mathcal{C}$ reduces to
\begin{equation}
\mathcal{C}=\frac{\left( z+n-1\right) C_{1}}{{l}^{z+1}}.
\end{equation}

It is worthwhile mentioning that although there are two constants $C_{1}$
and $C_{2}$ in the solutions (\ref{Solinf}), at infinity only the constant $%
C_{1}$, which is the geometrical mass of the black hole as we will show in
Sec. \ref{Therm}, appears in the conserved quantities along the radial
coordinate. Thus, one can conclude that our solution respects the no-hair
theorem.

\section{FINITE ACTION FOR GB-LIFSHITZ SOLUTIONS \label{Finite}}

The action (\ref{Act1}) is neither well defined nor finite. In order to get
a finite and well-defined action, one may add a few covariant boundary terms
to the action. The boundary term $I_{\mathrm{bdy}}$ is the sum of the
boundary terms which are needed to have a well-defined variational principle
and the counterterms which guarantee the finiteness of the action. $I_{%
\mathrm{bdy}}$, for the case of zero curvature boundary which is our
interest can be written as
\begin{equation}
I_{\mathrm{bdy}}=\frac{1}{8\pi }\int_{\partial \mathcal{M}}d^{n}x\sqrt{-h}%
\Big\{\Theta +2\alpha J-\frac{(n-1)(l^{2}-2\tilde{\alpha})}{l^{3}}+\frac{1}{2%
}f(e^{-4\lambda \Phi /(n-1)}U_{\gamma }U^{\gamma })\Big\}+I_{\mathrm{deriv}},
\label{Ibdy}
\end{equation}%
where the boundary $\partial \mathcal{M}$ is the hypersurface at some
constant $r$ and, therefore, the Greek indices take the values $0$ and $%
i=2...n $. In Eq. (\ref{Ibdy}), $h$ is the determinant of the induced metric $%
h_{\alpha \beta }$, $\Theta $ is the trace of the extrinsic curvature $%
\Theta _{\alpha \beta }$ and $J$ is the trace of \cite{DM2}
\begin{equation}
J_{\alpha \beta }=\frac{1}{3}(2\Theta \Theta _{\alpha \gamma }\Theta _{\beta
}^{\gamma }+\Theta _{\gamma \delta }\Theta ^{\gamma \delta }\Theta _{\alpha
\beta }-2\Theta _{\alpha \gamma }\Theta ^{\gamma \delta }\Theta _{\delta
\beta }-\Theta ^{2}\Theta _{\alpha \beta }).
\end{equation}%
For our case with the flat boundary, $I_{\mathrm{deriv}}$--which is a collection
of terms involving derivatives of the boundary field-- is zero. This is due
to the fact that both the curvature tensor constructed from the boundary
metric and covariant derivatives of $U_{\alpha }$ will not contribute to the
on-shell value of the action for the pure Lifshitz solution or its first
variation around the Lifshitz background. The boundary term for the matter
part of the action in the absence of the dilaton has been introduced in Ref.
\cite{Ross}. Here, we generalize it to the case of the Lifshitz solutions in the
presence of the dilaton field. For this case, we consider the matter part of the
boundary term to be a function of $e^{-4\lambda \Phi /(n-1)}U_{\gamma
}U^{\gamma }$ because\ it is constant on the boundary. One could find that $%
f(e^{-4\lambda \Phi /(n-1)}U_{\gamma }U^{\gamma })=a\left( -e^{-4\lambda
/(n-1)\Phi }U_{\gamma }U^{\gamma }\right) ^{1/2}$ where $a=4q/\left(
lb^{n-1}\right) $ [Note that $q$ can be substituted by (\ref{Constants})].
The variation of the total action $I_{\mathrm{tot}}=I_{\mathrm{bulk}}+I_{%
\mathrm{bdy}}$ about the solutions is
\begin{equation}
\delta I_{\mathrm{tot}}=\int d^{n}x\left( S_{\alpha \beta }\delta h^{\alpha
\beta }+S_{\alpha }^{L}\delta U^{\alpha }\right) ,
\end{equation}%
where
\begin{equation}
S_{\alpha \beta }=\frac{\sqrt{-h}}{16\pi }\left\{ \Pi _{\alpha \beta }-\frac{%
a}{2}e^{-2\lambda \Phi /(n-1)}\left( -U_{\gamma }U^{\gamma }\right)
^{-1/2}\left( U_{\alpha }U_{\beta }-U_{\gamma }U^{\gamma }h_{\alpha \beta
}\right) \right\} ,  \label{Sab}
\end{equation}%
\begin{equation}
S_{\beta }^{L}=-\frac{\sqrt{-h}}{16\pi }\left\{ 4e^{-4\lambda \Phi
/(n-1)}n^{\alpha }F_{\alpha \beta }+ae^{-2\lambda \Phi /(n-1)}\left(
-U_{\gamma }U^{\gamma }\right) ^{-1/2}U_{\beta }\right\} ,  \label{Sb}
\end{equation}%
with
\begin{equation}
\Pi _{\alpha \beta }=\Theta _{\alpha \beta }-\Theta h_{\alpha \beta
}+2\alpha (3J_{\alpha \beta }-Jh_{\alpha \beta })+\frac{(n-1)(l^{2}-2\tilde{%
\alpha})}{l^{3}}h_{\alpha \beta }.
\end{equation}%
In the Lifshitz background due to cancelation between different terms, $%
S_{\alpha \beta }=0$\ and $S_{\beta }^{L}=0$, and therefore the total action
satisfies $\delta I_{\mathrm{tot}}=0$ for arbitrary variations around
the Lifshitz solution. Thus, we have a finite on-shell action which defines a
well-defined variational principle for our background spacetime.

After constructing a well-defined finite action, one may compute the finite
stress tensor. This job has been done for asymptotically AdS spacetimes
which are dual to relativistic field theory \cite{Henn,Bal}. For
asymptotically Lifshitz spacetimes, the dual field theory is
nonrelativistic and, therefore, its stress tensor will not be covariant.
However one can define a stress tensor complex \cite{Ross}, consisting of
the energy density $\mathcal{E}{}$, energy flux ${}\mathcal{E}_{i}$,
momentum density ${}\mathcal{P}_{i}$, and spatial stress tensor $\mathcal{P}%
_{ij}$,
\begin{eqnarray}
\mathcal{E}{}{} &=&2S_{\ t}^{t}-S_{L}^{t}U_{t},\quad {}\mathcal{E}%
{}^{i}=2S_{\ t}^{i}-S_{L}^{i}U_{t},  \label{En} \\
{}\mathcal{P}_{i} &=&-2S_{\ i}^{t}+S_{L}^{t}U_{i},\quad \mathcal{P}%
_{i}^{j}=-2S_{\ i}^{j}+S_{L}^{j}U_{i},  \label{Pi}
\end{eqnarray}%
which satisfies the following conservation equations
\begin{equation}
\partial _{t}{}\mathcal{E}{}+\partial _{i}{}\mathcal{E}{}^{i}=0,\quad
\partial _{t}{}\mathcal{P}_{j}+\partial _{i}\mathcal{P}_{j}^{i}=0.
\label{Conserv}
\end{equation}%
In Eqs. (\ref{En}) and (\ref{Pi}), the Latin indices ($i,j$) go from $2$ to $%
n$ and $S_{\alpha \beta }$ and $S_{L}^{i}$ are given in Eqs. (\ref{Sab}) and
(\ref{Sb}).

\section{THERMODYNAMICS OF LIFSHITZ BLACK BRANES \label{Therm}}

Now, we are ready to consider the thermodynamics of Lifshitz black brane
solutions. The entropy in Gauss-Bonnet gravity can be calculated by using
\cite{meyers}%
\begin{equation}
\mathcal{S}=\frac{1}{4}\int d^{n-1}x\sqrt{\tilde{g}}\left( 1+2\alpha \tilde{R%
}\right) ,  \label{Entdef}
\end{equation}%
where $\tilde{g}$\ is the determinant of $\tilde{g}_{ij}$\ which is the
induced metric of the ($n-1$)-dimensional spacelike hypersurface of the Killing
horizon. Since we are dealing with a flat horizon, $\tilde{R}=0$ and, therefore,
the entropy per unit volume is
\begin{equation}
\mathcal{S}=\frac{r_{0}^{n-1}}{4}.  \label{Ent}
\end{equation}%
The temperature of the event horizon is given by%
\begin{equation}
T=\frac{1}{2\pi }\left( -\frac{1}{2}\nabla _{b}\chi _{a}\nabla ^{b}\chi
^{a}\right) _{r=r_{0}}^{1/2}  \label{T}
\end{equation}%
where $\chi =\partial _{t}$ is the Killing vector. Using (\ref{T}) and the
expansions of the metric functions near event horizon given in Sec. \ref%
{Cons}, one can obtain the temperature as
\begin{equation}
T=\frac{r_{0}^{z+1}}{4\pi l^{z+1}}\left( f^{\prime }g^{\prime }\right)
_{r=r_{0}}^{1/2}=\frac{r_{0}^{z+1}}{4\pi l^{z+1}}\sqrt{f_{1}g_{1}}.
\label{Temp}
\end{equation}

The conserved quantities of our solution can be calculated through the use
of the counterterm method of the previous section. The energy density of the
black brane can be calculated by using Eq. (\ref{En}) as
\begin{equation}
\mathcal{E}={\frac{\left( n-1\right) \sqrt{f}\left( 1-\sqrt{g}\right) {r}%
^{z+n-1}}{8\pi \,{l}^{z+1}}+}\frac{\left( n-1\right) \tilde{\alpha}\sqrt{f}%
\left( g^{3/2}-1\right) {r}^{z+n-1}}{12\pi \,{l}^{z+3}}+\frac{q^{2}k\left( e{%
^{-{2\lambda \,\Phi /(n-1)}}}\left( \frac{r}{b}\right) ^{n-1}-1\right) }{4{%
\pi l}^{z+1}}.
\end{equation}%
Inserting the large $r$ expansions given in the previous section for the
metric function in the above equation, the energy density may be calculated
as:
\begin{equation}
\mathcal{E}={\frac{\left( n-1\right) \left( l^{2}-2\tilde{\alpha}\right) %
\left[ \left( z+n-2\right) \left( z+n-1\right) \left( l^{2}-2\tilde{\alpha}%
\right) +2\tilde{\alpha}\left( z-1\right) ^{2}\right] C_{1}}{16\pi \mathcal{G%
}{l}^{z+3}}}.  \label{EnA}
\end{equation}%
For $\alpha =0$, $\mathcal{E}$ reduces to
\begin{equation}
\mathcal{E}={\frac{\left( n-1\right) C_{1}}{16\pi {l}^{z+1}}},
\end{equation}%
which is the energy of the spacetime obtained in Ref. \cite{Zang} as $%
\mathcal{E}=\left( n-1\right) m/16\pi {l}^{z+1}$, where $m$ was the
geometrical mass. This shows that in our solution, the constant $C_{1}$ is
the geometrical mass too. It is remarkable to note that by using (\ref{Pi})
one can calculate the angular momentum which is zero for our solution as one
expected.

Now, using Eqs. (\ref{Ent})-(\ref{EnA}), the constant $\mathcal{C}$ can be
written in terms of the thermodynamics quantities $T$, $\mathcal{S}$ and $%
\mathcal{E}$ as
\begin{equation*}
\mathcal{C}=16\pi T\mathcal{S}=\frac{16\pi \left( n+z-1\right) \mathcal{E}}{%
n-1}.
\end{equation*}%
Thus, one obtains
\begin{equation}
\mathcal{E}=\left( \frac{n-1}{n+z-1}\right) T\mathcal{S}.  \label{ETS}
\end{equation}

Finally, we would like to perform thermal stability analysis in the case of
asymptotically Lifshitz solution in dilaton Gauss-Bonnet gravity. Since our
solution is uncharged, the positivity of the heat capacity $C=T/(dT/d%
\mathcal{S})$ is sufficient to ensure the local stability. In order to
calculate the heat capacity, we first use the first law of thermodynamics $d%
\mathcal{E}=Td\mathcal{S}$ with the relation (\ref{ETS}) for the energy
density to obtain%
\begin{equation}
\frac{dT}{d\mathcal{S}}=\frac{z}{n-1}\frac{T}{\mathcal{S}}.  \label{logT}
\end{equation}
Thus, the heat capacity can be obtained as%
\begin{equation*}
C=\frac{(n-1)}{z}\mathcal{S}=\frac{(n-1)r_{0}^{n-1}}{4z}
\end{equation*}
which is positive and therefore our black brane solution is thermally
stable. Also, it is worth noting that the curve of $\log T$ versus $\log
\mathcal{S}$ is a line with slope $z/(n-1)$:%
\begin{equation*}
\log T=\frac{z}{n-1}\log \mathcal{S}+\Gamma ,
\end{equation*}%
where $\Gamma $ is an integration constant.

\section{CONCLUSION}

In this paper, we considered asymptotic Lifshitz black branes of the
effective supergravity action coming from superstrings, which contains the GB
term and a dilaton field, in the presence of a massless gauge field.
Although it is known that the GB term is coupled to the dilaton, we
considered them decoupled for simplicity. By variation of the action, we
found four independent equations of motion. Then, we fixed the parameters of
our model such that the asymptotic Lifshitz behavior is supported. Next, we
obtained two independent constants along the radial coordinate by combining
the equations of motion. We combined these two constants in order to get a
constant $\mathcal{C}$ which was proportional to the energy density.\ In
addition, we calculated the value of this constant quantity at the horizon
in terms of the thermodynamic quantities, temperature and entropy. Also using
the large-$r$ behaviors of metric functions, we found the value of this
constant at large $r$. Although there are two independent constants in our
theory, we found out that only one will appear at infinity. This shows that
our solution respects the no-hair theorem. In order to compute the finite stress
energy tensor, we constructed the well-defined finite action. This action is
the generalization of the action presented in the case where no dilaton
field exists \cite{Ross}. By calculating the value of conserved quantity
(energy density) in terms of the constant $\mathcal{C}$, we obtained the
relation between energy density, temperature and entropy. This relation is
the generalization of the well-known Smarr formula for AdS black holes.
Finally, we performed the thermal stability analysis on our solution and
showed that our black brane solution is stable under thermal perturbations.

In this paper we studied the thermodynamics of uncharged dilaton Lifshitz
black branes in the context of GB gravity where the GB term is decoupled
from the dilaton. This work can be extended in various ways. First, one
can consider the case where the GB term is coupled to the dilaton field.
Second, one may seek the thermodynamics of the linearly and nonlinearly
charged Lifshitz black branes of this theory. Another interesting case is
the consideration of black holes that their horizons' geometries are not
flat. A study of these solutions can also be extended to the case of effective
supergravity action coming from superstrings, which contains higher-curvature
terms. We hope to address Some of the above-mentioned suggestions in future works.

\section*{ACKNOWLEDGEMENTS}

We thank the Shiraz University Research Council. This work has been supported
financially by the Research Institute for Astronomy and Astrophysics of Maragha,
Iran.


\begin{thebibliography}{99}
\bibitem{Str} D. J. Gross and J. S. Loan, "The quartic effective action for
the heterotic string", Nucl. Phys. B \textbf{291}, 41 (1987);\newline
M. C. Bento and O. Bertolami, "Cosmological solutions of higher-curvature
string effective theories with dilatons", Phys. Lett. B \textbf{368, }198
(1996).

\bibitem{Char} C. Charmousis, B. Gout\'{e}raux and E. Kiritsis,
"Higher-derivative scalar-vector-tensor theories: black holes, Galileons,
singularity cloaking and holography", JHEP \textbf{1209, }011 (2012).

\bibitem{Ohta1} K. Maeda, N. Ohta and Y. Sasagawa, "AdS black hole solution
in dilatonic Einstein-Gauss-Bonnet gravity", Phys. Rev. D \textbf{83},
044051 (2011).

\bibitem{Ohta2} Z.~K.~Guo, N.~Ohta and T.~Torii, "Black holes in the
dilatonic Einstein-Gauss-Bonnet theory in various dimensions
I-Asymptotically flat black holes-", Prog. Theor. Phys. \textbf{120, }581
(2008); Z.~K.~Guo, N.~Ohta and T.~Torii, "Black holes in the dilatonic
Einstein-Gauss-Bonnet theory in various dimensions II-Asymptotically AdS
topological black holes-", Prog. Theor. Phys. \textbf{121, }253 (2009);
N.~Ohta and T.~Torii, "Black holes in the dilatonic Einstein-Gauss-Bonnet
theory in various dimensions III-Asymptotically AdS black holes with $k=\pm 1
$-", Prog. Theor. Phys. \textbf{121}, 959 (2009); N.~Ohta and T.~Torii,
"Black holes in the dilatonic Einstein-Gauss-Bonnet theory in various
dimensions IV-Topological black holes with and without cosmological term-",
Prog. Theor. Phys. \textbf{122, }1477 (2009); K.~Maeda, N.~Ohta and
Y.~Sasagawa, "Black hole solutions in string theory with Gauss-Bonnet
curvature correction", Phys. Rev. D \textbf{80,} 104032 (2009); C.~M.~Chen,
D.~V.~Gal'tsov, N.~Ohta and D.~G.~Orlov, "Global solutions for
higher-dimensional stretched small black holes", Phys. Rev. D \textbf{81},
024002 (2010); N.~Ohta and T.~Torii, "Global structure of black holes in
string theory with Gauss-Bonnet correction in various dimensions", Prog.
Theor. Phys. \textbf{124,} 207 (2010); N.~Ohta and T.~Torii, "Charged black
holes in string theory with Gauss-Bonnet correction in various dimensions",
Phys. Rev. D \textbf{86}, 104016 (2012).

\bibitem{Mald} J. M. Maldacena, "The large N limit of superconformal field
theories and supergravity", Adv. Theor. Math. Phys. \textbf{2, }231 (1998);%
\newline
E. Witten, "Anti de Sitter space and holography", Adv. Theor. Math. Phys.
\textbf{2, }253 (1998);\newline
E. Witten, "Anti-de Sitter Space, Thermal phase transition, and confinement
in gauge theories", Adv. Theor. Math. Phys. \textbf{2, }505 (1998).

\bibitem{Lif} S. Kachru, X. Liu and M. Mulligan, "Gravity duals of
Lifshitz-like fixed points", Phys. Rev. D \textbf{78, }106005 (2008).

\bibitem{manal2} A. Alvarez, E. Ayon-Beato, H. A. Gonzalez and M. Hassaine,
"Nonlinearly charged Lifshitz black holes for any exponent $z>1$", JHEP
\textbf{06, }041 (2014).

\bibitem{manal1} M. H. Dehghani, R. Pourhasan and R. B. Mann, "Charged
Lifshitz black holes", Phys. Rev. D \textbf{84, }046002 (2011).

\bibitem{cmet} M. H. Dehghani, Ch. Shakuri and M. H. Vahidinia, "Lifshitz
black brane thermodynamics in the presence of a nonlinear electromagnetic
field", Phys. Rev. D \textbf{87, }084013 (2013).

\bibitem{cmlt} M. H. Dehghani and R. B. Mann, "Thermodynamics of
Lovelock-Lifshitz black branes", Phys. Rev. D \textbf{82, }064019 (2010);%
\newline
M. H. Dehghani and Sh. Asnafi, "Thermodynamics of rotating Lovelock-Lifshitz
black branes", Phys. Rev. D \textbf{84, }064038 (2011).

\bibitem{cmlnt} M. H. Dehghani and R. B. Mann, "Lovelock-Lifshitz Black
Holes", JHEP \textbf{1007, }019 (2010);\newline
W. G. Brenna, M. H. Dehghani and R. B. Mann, "Quasi-topological Lifshitz
black holes", Phys. Rev. D \textbf{84, }024012 (2011).

\bibitem{hcc} E. Ayon-Beato, A. Garbarz, G. Giribet and M. Hassaine,
"Lifshitz black hole in three dimensions", Phys. Rev. D \textbf{80, }104029
(2009);\newline
R. G. Cai, Y. Liu and Y. W. Sun, "A Lifshitz Black Hole in Four Dimensional $%
R^{2}$ Gravity", JHEP \textbf{0910, }080 (2009);\newline
E. Ayon-Beato, A. Garbarz, G. Giribet and M. Hassaine, "Analytic Lifshitz
black holes in higher dimensions", JHEP \textbf{1004, }030 (2010);\newline
M. Bravo-Gaete and M. Hassaine, "Thermodynamics of charged Lifshitz black
holes with quadratic corrections", Phys. Rev. D \textbf{91,} 064038 (2015).

\bibitem{dcanal} J. Tarrio and S. Vandoren, "Black holes and black branes in
Lifshitz spacetimes", JHEP \textbf{1109,} 017 (2011);\newline
M. H. Dehghani, A. Sheykhi and S. E. Sadati, "Thermodynamics of nonlinear
charged Lifshitz black branes with hyperscaling violation", Phys. Rev. D
\textbf{91}, 124073 (2015).

\bibitem{Zang} M. Kord Zangeneh, A. Sheykhi and M. H. Dehghani,
"Thermodynamics of topological nonlinear charged Lifshitz black holes",
Phys. Rev. D \textbf{92}, 024050 (2015).

\bibitem{peet} G. Bertoldi, B. A. Burrington and A. W. Peet, "Thermal
behavior of charged dilatonic black branes in AdS and UV completions of
Lifshitz-like geometries", Phys. Rev. D \textbf{82, }106013 (2010).

\bibitem{Ross} S. F. Ross and O. Saremi, "Holographic stress tensor for
nonrelativistic theories", JHEP \textbf{09, }009 (2009).

\bibitem{Ayz} D. Ayzenberg and N. Yunes, "Slowly-rotating black holes in
Einstein-Dilaton-Gauss-Bonnet gravity: Quadratic order in spin solutions",
Phys. Rev. D \textbf{90}, 044066 (2014).

\bibitem{STR} D. G. Boulware and S. Deser, "String-generated gravity
models", Phys. Rev. Lett. \textbf{55, }2656 (1985).

\bibitem{DM2} M. H. Dehghani and R. B. Mann, "Thermodynamics of rotating
charged black branes in third order lovelock gravity and the counterterm
method", Phys. Rev. D \textbf{73, }104003 (2006); M. H. Dehghani, N. Bostani
and A. Sheykhi, "Counterterm method in Lovelock theory and horizonless
solutions in dimensionally continued gravity", ibid. \textbf{73, }104013
(2006).

\bibitem{Henn} M. Henningson and K. Skenderis, "The holographic Weyl
anomaly", JHEP \textbf{07, }023 (1998).

\bibitem{Bal} V. Balasubramanian and P. Kraus, "A stress tensor for anti-de
Sitter gravity", Commun. Math. Phys. \textbf{208, }413 (1999).

\bibitem{meyers} T. Jacobson and R. C. Myers, "Entropy of Lovelock black
holes", Phys. Rev. Lett. \textbf{70}, 3684 (1993).
\end{thebibliography}
\end{document}